\documentclass[a4paper]{spie}  

\usepackage[utf8]{inputenc}
\usepackage{amsmath,amsfonts,amssymb}
\usepackage{graphicx}
\usepackage[colorlinks=true, allcolors=blue]{hyperref}
\usepackage{upgreek}
\usepackage{epstopdf}
\usepackage{epsfig}
\usepackage{units}

\usepackage{acronym}
\usepackage[caption=false]{subfig}

\acrodef{AO}[AO]{adaptive optics}
\acrodef{CCD}{charged-coupled device}
\acrodef{DARC}{Durham Adaptive Optics Real-Time Controller}
\acrodef{DM}[DM]{deformable mirror}
\acrodef{ELT}[ELT]{Extremely Large Telescope}
\acrodef{ExAO}[ExAO]{extreme adaptive optics}
\acrodef{FLAO}{First Light AO}
\acrodef{FoV}{field of view}
\acrodef{ISYS}[ISYS]{Institute for System Dynamics Stuttgart}
\acrodef{KIT}[KIT]{Karlsruhe Institute for Technology}
\acrodef{KOOL}[KOOL]{Koenigstuhl Observatory Opto-Mechatronics Laboratory}
\acrodef{LBT}[LBT]{Large Binocular Telescope}
\acrodef{LBTI}[LBTI]{Large Binocular Telescope Interferometer}
\acrodef{LBTIAO}{\ac{LBTI} AO}
\acrodef{LSW}[LSW]{Landessternwarte}
\acrodef{MCF}[MCF]{multi-core fiber}
\acrodef{MFD}[MFD]{mode-field diameter}
\acrodef{MLA}[MLA]{micro-lens array}
\acrodef{MLR}[MLR]{micro-lens ring}
\acrodef{MM}[MM]{multi-mode}
\acrodef{MMF}[MMF]{multi-mode fiber}
\acrodef{MPIA}[MPIA]{Max Plank Institute for Astronomy}
\acrodef{NA}[NA]{numerical aperture}
\acrodef{NCP}[NCP]{non-common path}
\acrodef{NAIR}[NAIR]{Novel Astronomical Instrumentation based on photonic light Reformating}
\acrodef{NIR}[NIR]{near-infrared}
\acrodef{PSF}[PSF]{point-spread function}
\acrodef{PSD}[PSD]{power spectral density}
\acrodef{SM}[SM]{single-mode}
\acrodef{SMF}[SMF]{single-mode fiber}
\acrodef{SR}[SR]{Strehl rati}
\acrodef{WFS}[WFS]{wavefront sensor}

\newcommand{\mum}{\upmu \mathrm{m}}

\title{Focal Plane Tip-Tilt Sensing for Improved Single-Mode Fiber Coupling using a 3D-printed Microlens-Ring}

\author[a]{Philipp Hottinger}
\author[a]{Robert J. Harris}
\author[b,c,d]{Philipp-Immanuel Dietrich}
\author[b,c]{Matthias Blaicher}
\author[f]{Andrew Bechter}
\author[f]{Jonathan Crass}
\author[e]{Martin Glück}
\author[g]{Jörg-Uwe Pott}
\author[h]{Nazim A. Bharmal}
\author[h]{Alastair Basden}
\author[h]{Tim J. Morris}
\author[b,c,d]{Christian Koos}
\author[e]{Oliver Sawodny}
\author[a]{Andreas Quirrenbach}

\affil[a]{Landessternwarte~(LSW), Zentrum für Astronomie der Universität Heidelberg, Königstuhl~12, 69117~Heidelberg, Germany}
\affil[b]{Institute of Microstructure Technology~(IMT), Karlsruhe Institute of Technology~(KIT), Hermann-von-Helmholtz-Platz~1, 76344~Eggenstein-Leopoldshafen, Germany}
\affil[c]{Institute of Photonics and Quantum Electronics~(IPQ), Karlsruhe Institute of Technology~(KIT), Engesserstr.~5, 76131~Karlsruhe, Germany}
\affil[d]{Vanguard Photonics GmbH, Hermann-von-Helmholtz-Platz 1, 76344~Eggenstein-Leopoldshafen, Germany}
\affil[e]{Institute for System Dynamics, University of Stuttgart, Waldburgstr. 19, 70563 Stuttgart, Germany}
\affil[f]{Department of Physics, University of Notre Dame, 225~Nieuwland Science Hall, Notre Dame, IN~46556, USA}
\affil[g]{Max-Planck-Institute for Astronomy, Königstuhl~17, 69117~Heidelberg, Germany}
\affil[h]{Durham University, Centre for Advanced Instrumentation, Department of Phyiscs, South Road, Durham, UK, DH1 3LE}

\authorinfo{Further author information: (Send correspondence to P.H.)\\P.H.: E-mail: phottinger@lsw.uni-heidelberg.de, Telephone: +49 6221 54 1727}

\begin{document}
\maketitle

\begin{abstract}

Modern \acf{ExAO} systems achieving diffraction-limited performance open up new possibilities for instrumentation. Especially important for the fields of spectroscopy and interferometry is that it opens the possibility to couple light into \acfp{SMF}. However, due to their small size, efficient coupling is very sensitive to the quality of the fiber alignment, beam drifts and higher-frequency tip-tilt aberrations caused by telescope mechanics and vibrations. These residual aberrations are not always sensed and corrected by the \ac{AO} system, leading to unacceptable losses. This is particularly severe for the Extremely Large Telescopes, where their huge structure will mean vibrations increase and optimal \ac{AO} solutions are even more difficult to implement.

We have created a focal plane sensor to correct for residual aberrations by surrounding the SMF with six Multi-mode fibers (MMFs).  On each of the MMFs sits a printed freeform lens, making up a six-element \acf{MLR} to refract the light into these surrounding MMFs and thus minimizing light loss in the gap between the fiber cores. This means when the beam is near diffraction limited and centered almost all light couples to the \ac{SMF}. When the beam is misaligned, it couples to the surrounding cores, which are read out by a detector and processed by the Durham Adaptive Optics Real-Time Control (DARC) software driving a tip-tilt mirror. Currently we are aiming to detect and correct only tip-tilt aberrations. However, choosing to surround the central fiber with six sensing locations potentially allows us to investigate higher order correction modes.

Here we present the design and performance our prototype system. This has been designed for use with the iLocater fiber injection system at the Large Binocular Telescope and can easily be scaled to larger telescopes. We present test results from the KOOL laboratory in Heidelberg and initial integration with the iLocater instrument.

\end{abstract}


\section{Introduction}
\label{sec:introduction}
\acresetall 

In recent years, the development of advanced \ac{AO} systems has opened up new possibilities: ever-improved image quality, a wider \ac{FoV} and greater sky coverage, leading to new discoveries in fields of astronomy from direct imaging of exoplanets \cite{Rodigas2014} to examining the motions of the stars around Sgr A* \cite{gillessen2009} .
In particular the development of \ac{ExAO} has enabled 8-10m class telescopes to achieve diffraction-limited optical performance, where they would otherwise be seeing-limited \cite{Esposito2011, macintosh2014,Beuzit2008}. Whilst the \ac{FoV} is limited, this enhanced capability has allowed objects to be viewed with far more detail than ever before.



This improvement has also opened new doors in the field of spectroscopy. Conventionally, spectrographs had large entrance apertures, matched to the seeing limit. In the case of fiber fed spectrographs, this meant using \acp{MMF}, with core diameters on the order of 100 microns. Due to their large core diameter, the light from the telescope could be efficiently coupled from the telescope, though this formed a large entrance slit to the spectrograph.  Due to the instrument scaling laws, the instruments behind had to be appropriately scaled in size \cite{lee2000}. Using an \ac{ExAO} system reduces the size of the \ac{PSF} to the diffraction limit, which in turn allows the size of the fiber core to be reduced. These fibers are called \acp{SMF} and by coupling into their smaller entrance aperture and usually smaller \ac{NA}, the spectrographs that are fed with \acp{SMF} can not only be reduced in size, while maintaining the same spectral resolution, but are also free of conventional modal noise \cite{Crepp2016}. This leads to increased stability and reduced cost for the instrument.
Whilst in principle this is an ideal solution, it comes at the cost of increased alignment tolerances. In recent years, there have been several attempts to couple light from large telescope into \acp{SMF}, but the coupling efficiency highly depends on initial fiber alignment, as well as long term movement (beam drifts) and higher-frequency tip-tilt motions \cite{Bechter2016}. These can originate both from residual atmospheric aberrations or from telescope and instrument flexure and vibrations.



We have developed a fiber based focal plane tip-tilt sensor to compensate for the movements and tip-tilt vibrations.
The concept was introduced in 2017 \cite{Dietrich2017} and a modified preliminary design was presented in 2018, as a tip-tilt sensing \ac{MLA} \cite{Hottinger2018}.\footnote{
	Despite the difference in name, the working principles presented in this work are identical to the ones in Ref.~\citenum{Hottinger2018}.
	The actual geometry of the lenses has inspired us to rename the 3D printed lens to \acf{MLR} instead of \acf{MLA}, due to design having a central aperture and the overall shape being point symmetric.
	The fiber arrangement has changed slightly as printing restrictions required the use of \acp{MMF} with larger core sizes.}
The tip-tilt sensor is based upon a fiber bundle consisting of the ``science" \ac{SMF} and six surrounding sensing \acp{MMF}.
A \ac{MLR} sits on top of the fiber bundle tip and refracts increasing amounts of light into the corresponding surrounding sensing fibers as the beam gets misaligned.
Analysis of the amount of light coupled into these fibers then allows reconstruction of the actual beam position, i.e. the tip-tilt.
This device is designed to be retrofitted to any \ac{SMF} fed spectrograph, but our prototype is specifically designed for the prototype of the iLocater spectrograph\cite{Crepp2016} at the \ac{LBT}.

In this work we outline the working principles of the fiber-based tip-tilt sensor and the corresponding application (Sec.~\ref{sec:design}), as well as its performance in lab conditions (Sec.~\ref{sec:results}).
In Sec.~\ref{sec:discussion} we disucss advantages of this sensor before summarizing and highlighting future work in Sec.~\ref{sec:conclusion}.

\section{Design}
\label{sec:design}

\begin{figure}
	\begin{center}
		\begin{tabular}{c c c}
			\subfloat[Centered]{
        \includegraphics[width=.3\textwidth]{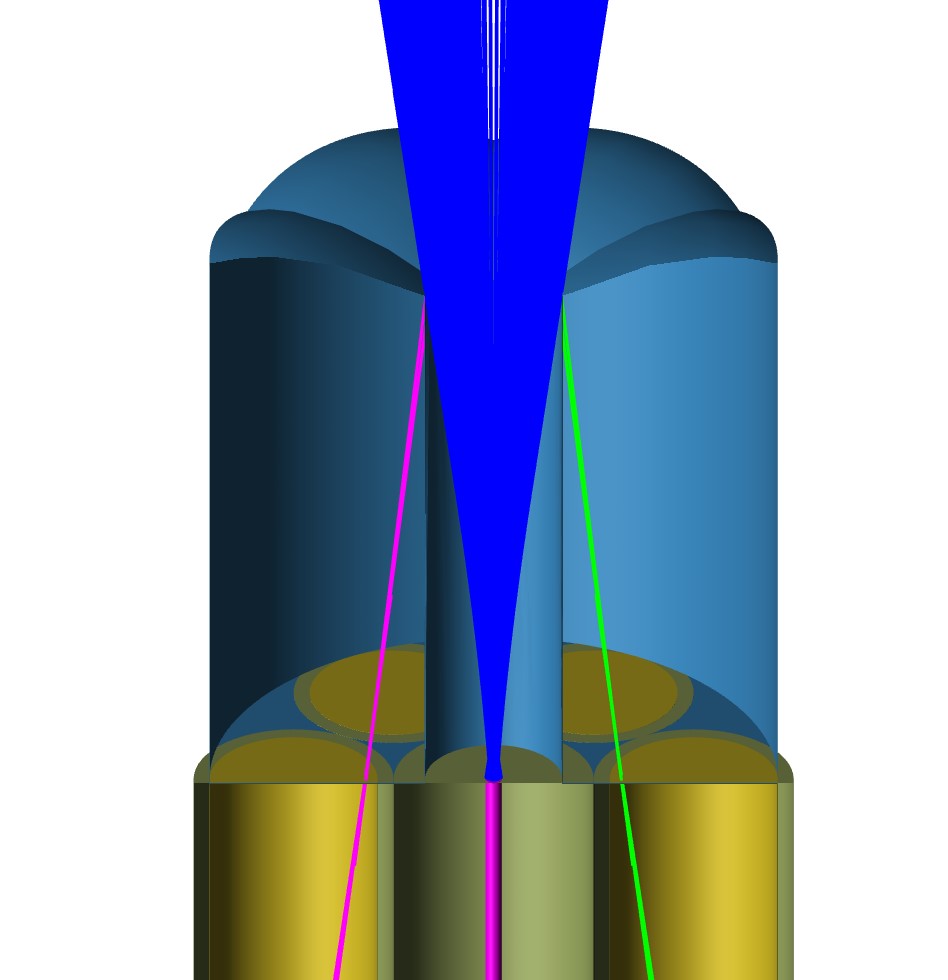}
        \label{fig:shaded_0mu}
      } &
			\subfloat[Offset $5\mum$]{
        \includegraphics[width=.3\textwidth]{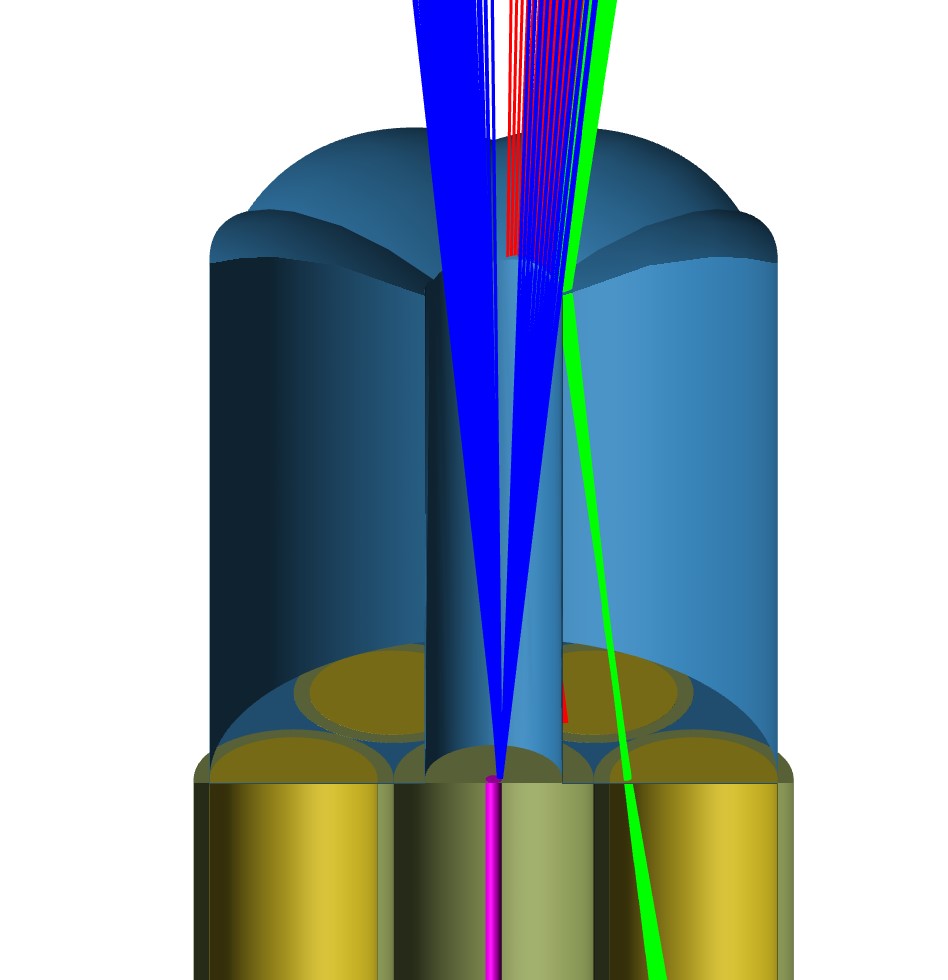}
        \label{fig:shaded_5mu}
      } &
			\subfloat[Offset $10\mum$]{
        \includegraphics[width=.3\textwidth]{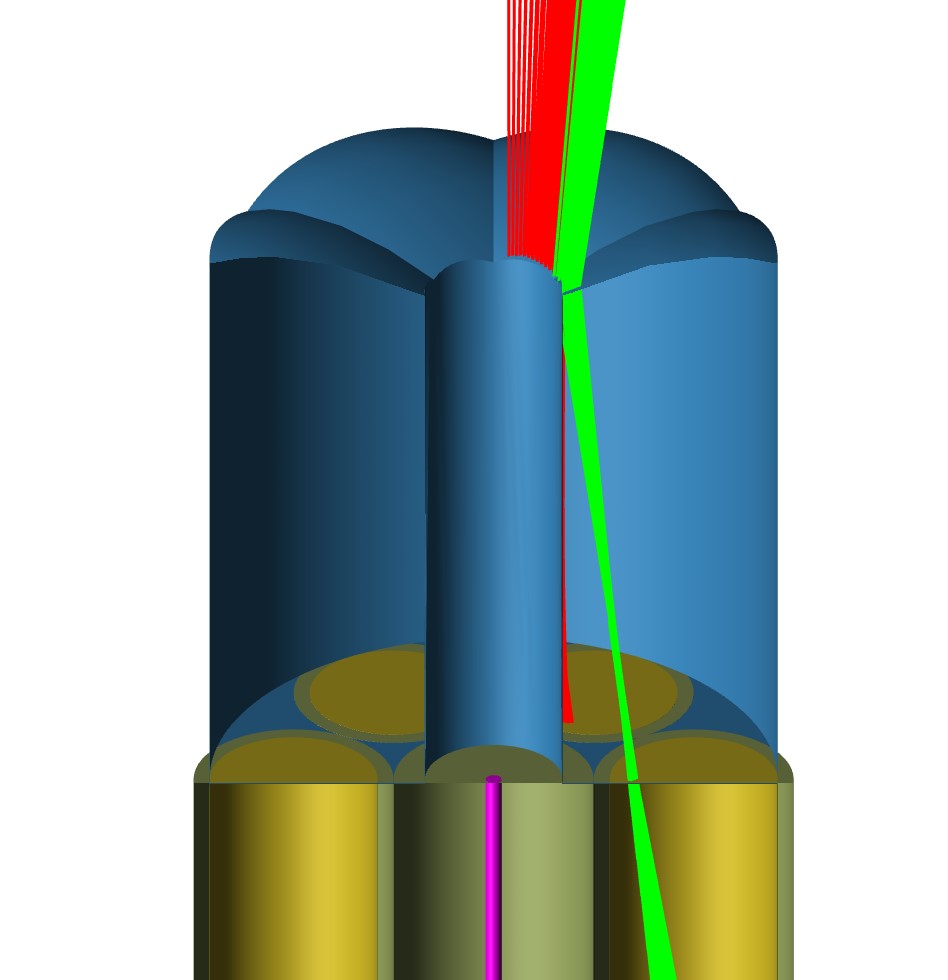}
        \label{fig:shaded_10mu}
      }
		\end{tabular}
	\end{center}
	\caption{
		Modeled ray propagation through the \acf{MLR} for an incoming beam that is \textbf{(a)} aligned and offset by \textbf{(b)} $5 \mum$ and \textbf{(b)} $10\mum$, respectively. In the platescale of iLocater frontend (at 1 $\mum$), the diffraction limit $1.22\lambda/D \sim 60\mathrm{mas}$  corresponds to $\sim$ 3.9$\mum$ offset. Please note that the number of rays propagating to the tip of the \acf{SMF} does not correspond to the coupling efficiency.}
	\label{fig:shaded}
\end{figure}

\subsection{Optical principle}

At the focal plane of the telescope, the tip-tilt sensor is formed of a fiber bundle, which consists of one central \ac{SMF} (Fibercore SM980, $1/e^2$ \ac{MFD}=5.8$\mum$) and six surrounding \acp{MMF} (Thorlabs, core size 105$\mum$, NA=0.22).
The central science fiber guides the light from the telescope to the spectrograph. This fiber is taken from the same production batch as the fiber for iLocater, which allows us to match the \ac{MFD} and therefore increase throughput.
The surrounding \ac{MMF} are fed to the sensing system. As these fibers do not feed the spectrograph we can make use of the larger core diameter \acp{MMF}, to allow for better coupling through reduced alignment tolerances. A small fraction of the light is coupled into these fibers even when the \ac{PSF} is on axis to allow for correction feedback.
This principle is illustrated in Fig.~\ref{fig:shaded} for an aligned beam (a) that achieves maximum coupling efficiency into the science fiber while only a low amount of light is evenly distributed into the surrounding sensing fibers.
When the incoming beam is misaligned by $5 \mum$ (b) and $10 \mum$ (c), the amount of light that couples to the sensing fibers located in the direction of the displacement increases.



\subsection{Manufacturing}

\begin{figure}
	\begin{center}
        \includegraphics[width=.8\textwidth]{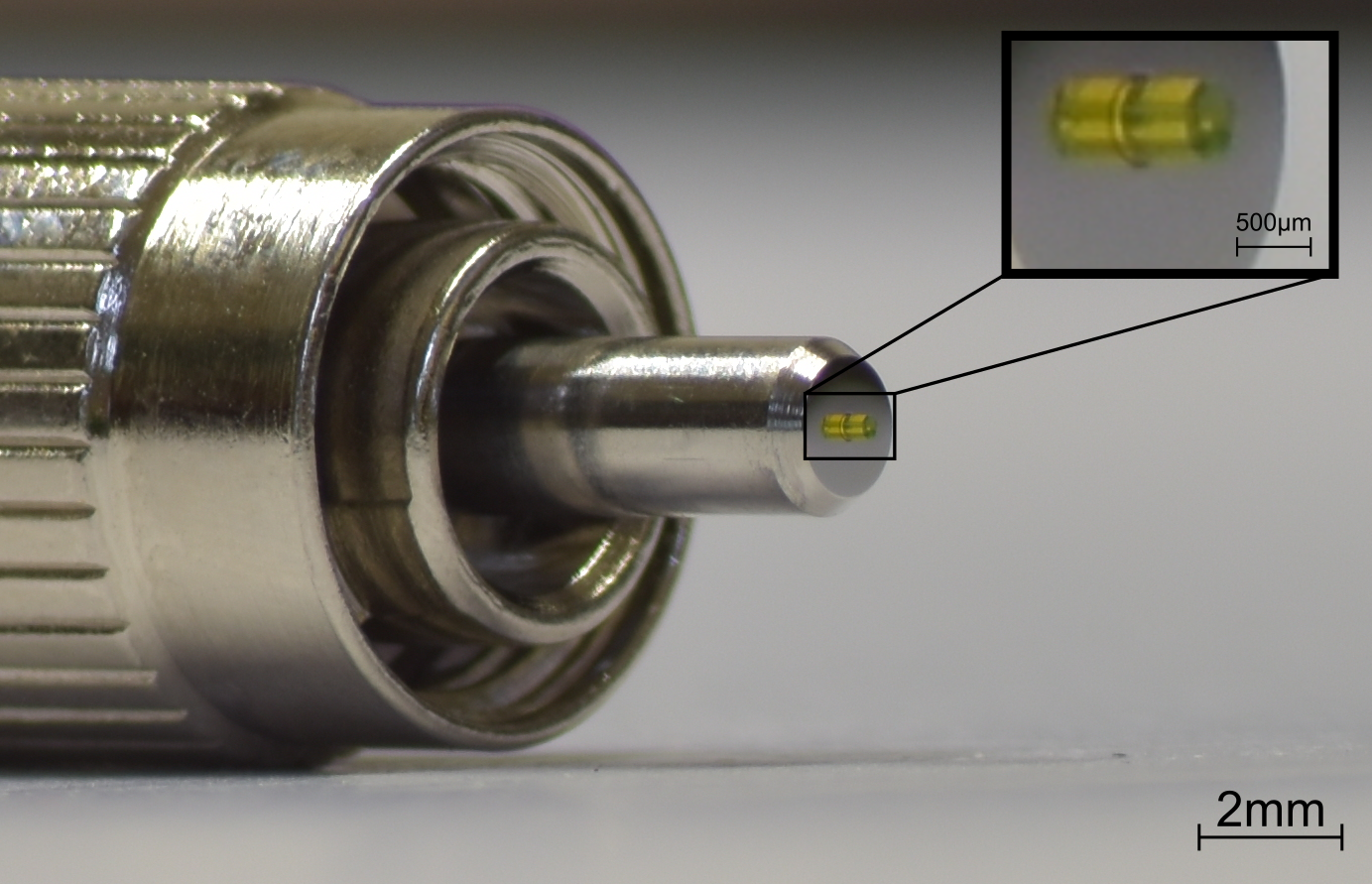}
	\end{center}
	\caption{
  DSLR photograph of \acf{MLR}.
	The lens stands around 380 $\mum$ tall and around 400 $\mum$ in diameter.
  }
	\label{fig:mlr-real}
\end{figure}

As the design for our \ac{MLR} is unusual, this would be excessively expensive to design and produce using conventional methods. To make our sensor economically viable we use an in-situ printing technique developed for the telecommunications industry and recently tested for astronomical applications. \cite{Dietrich2017,Dietrich2018}
This technique uses 3D-lithography by two-photon polymerization of a commercial IP-resist from \textit{nanoscribe} and allows us to print directly on the tip of the fiber.
Printing on the tip of the fiber allows very precise alignment of the lenses to the cores, as the position of the individual cores is measured before printing and the printing position adjusted to compensate for any differences between design and manufactured bundle.


Fig.~\ref{fig:mlr-real} shows the completed \ac{MLR} on top of a FC/PC connector ferrule.
The lens stands approximately 380 $\mum$ tall and has a diameter of approximately 400 $\mum$.
The central aperture has a diameter of approximately 80 $\mum$ leaving the light path to the science fiber uneffected. Using an aperture instead of a lens means reflections and surface quality do not play a role in the \ac{SMF} coupling and the iLocater system does not have to be modified to accommodate the new lens.
There is a limited effect due to the edges of this hole vignetting the beam, which results in a slight chromatic coupling efficiency difference.

\subsection{Correction}

The six surrounding \acp{MMF} are separated from the \ac{SMF} using a 3D printed fiber breakout and rearranged to form a linear array, which is then re-imaged onto a InGaAs camera (First Light C-Red 2).
The fluxes of the individual fibers are read, and processed by \ac{DARC} \cite{basden2010,basden2012}, running on a computer equipped with a consumer grade CPU (i5-8400).
\ac{DARC} then reconstructs the actual centroid position from the six fluxes using a sine-fit approach with some calibration correction.
The loop is then closed by an integration correction, feeding a signal to a tip-tilt mirror upstream.

\section{Results}
\label{sec:results}

\begin{figure}
	\begin{center}
		\begin{tabular}{c c}
			\subfloat[Modeled]{
        \includegraphics[width=.4\textwidth]{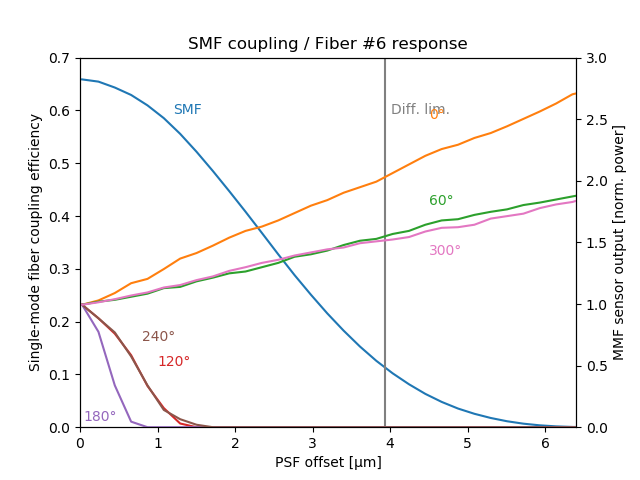}
        \label{fig:response-theory}
      } &
			\subfloat[Measured]{
        \includegraphics[width=.4\textwidth]{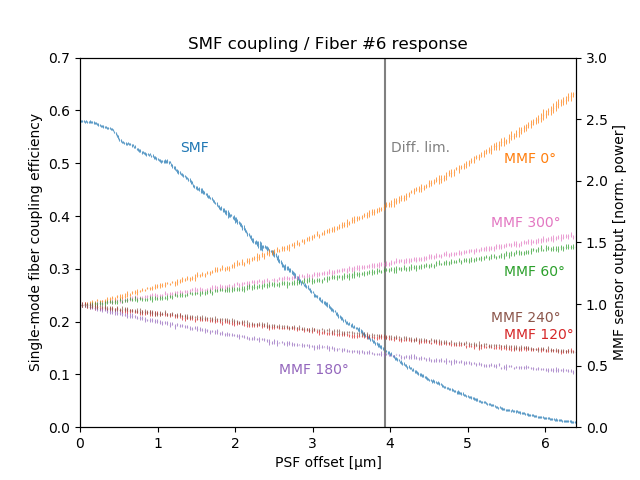}
        \label{fig:response-lab}
      }
		\end{tabular}
	\end{center}
	\caption{
 		Fiber response in respect to centroid position for both modeled throughput in the ray tracing software \textit{Zemax} (a) and for measured throughput at the iLocater fronted prototype in the lab (b).
		The vertical gray line denotes the diffraction limit at $\lambda \approx 1 \mum$.
		\ac{SMF} coupling (light blue markers, left y-axis)
		\acp{MMF} coupling (right y-axis) for all six sensing fibers, from same direction as the misalignment (orange marker, corresponding to very right fiber with green rays in Fig.~\ref{fig:shaded}), the two adjacent fibers (green, pink, corresponding to second fiber from right with red rays on Fig.~\ref{fig:shaded}), to the three fibers on the opposite direction (brown, red, violet, corresponding to two left fibers with pink rays on Fig.~\ref{fig:shaded}).
		All \acp{MMF} have differing throughputs, which are normalized in this graph for illustration.
		}
	\label{fig:response}
\end{figure}

Setup and optimization of the fiber-based tip-tilt sensor and the corresponding control system were carried out at the \ac{KOOL}\cite{Hottinger2018}, in Heidelberg, Germany.
Initial integration tests were conducted at the iLocater frontend prototype at the University of Notre-Dame in Indiana, USA.

Fig.~\ref{fig:response} shows the response for the seven individual fibers depending on the centroid position for both modeled throughput in the ray tracing software \textit{Zemax} (a) and for measured throughput at the iLocater fronted prototype in the lab (b).
As the incoming beam is de-centered, the \ac{SMF} coupling (light blue markers, left $y$-axis) decreases significantly within a few $\mum$.
On the right $y$-axis, the response on the sensing fibers is plotted.
The amount of coupled light increases for the \ac{MMF} corresponding to the direction of the offset (orange marker, corresponding to very right fiber with green rays in Fig.~\ref{fig:shaded}) as well as the two adjacent fibers (green, pink, corresponding to second fiber from right with red rays on Fig.~\ref{fig:shaded}).
The throughput of the \acp{MMF} opposite to the misalignment decreases (brown, red, violet, corresponding to two left fibers with pink rays on Fig.~\ref{fig:shaded}).
All \acp{MMF} have differing throughputs which are normalized in this graph for illustration.
The actual difference originates in residual aberrations in the \ac{PSF} of the optical setup, in the reconstruction algorithm this is accounted for by a calibration correction.

The overall flux in the six sensing fibers amounts to $2.3\%$ of the overall incoming flux in the lab measurements compared to $10\%$ expected from modeling, which is still being investigated.
The \ac{SMF} coupling efficiency is designed to amount to $67\%$ of the overall incoming light which is less than the $\sim80\%$ that is theoretically possible when coupling an Airy pattern into a \ac{SMF}. \cite{Shaklan1988}
The measured maximum coupling efficiency is  $58\%$ somewhat lower than the expected performance from modeling.
This $\sim10\%$ percentage points difference corresponds to the $70\%$ coupling efficiency into a regular bare \ac{SMF} that was achieved on the same setup which is also $\sim 10$ percentage points below the achievable maximum coupling efficiency.
We therefore account that difference to residual aberrations in the beam and likely induced in the optics used in generating a simulated telescope beam in laboratory testing.

\begin{figure}
	\begin{center}
        \includegraphics[width=.6\textwidth]{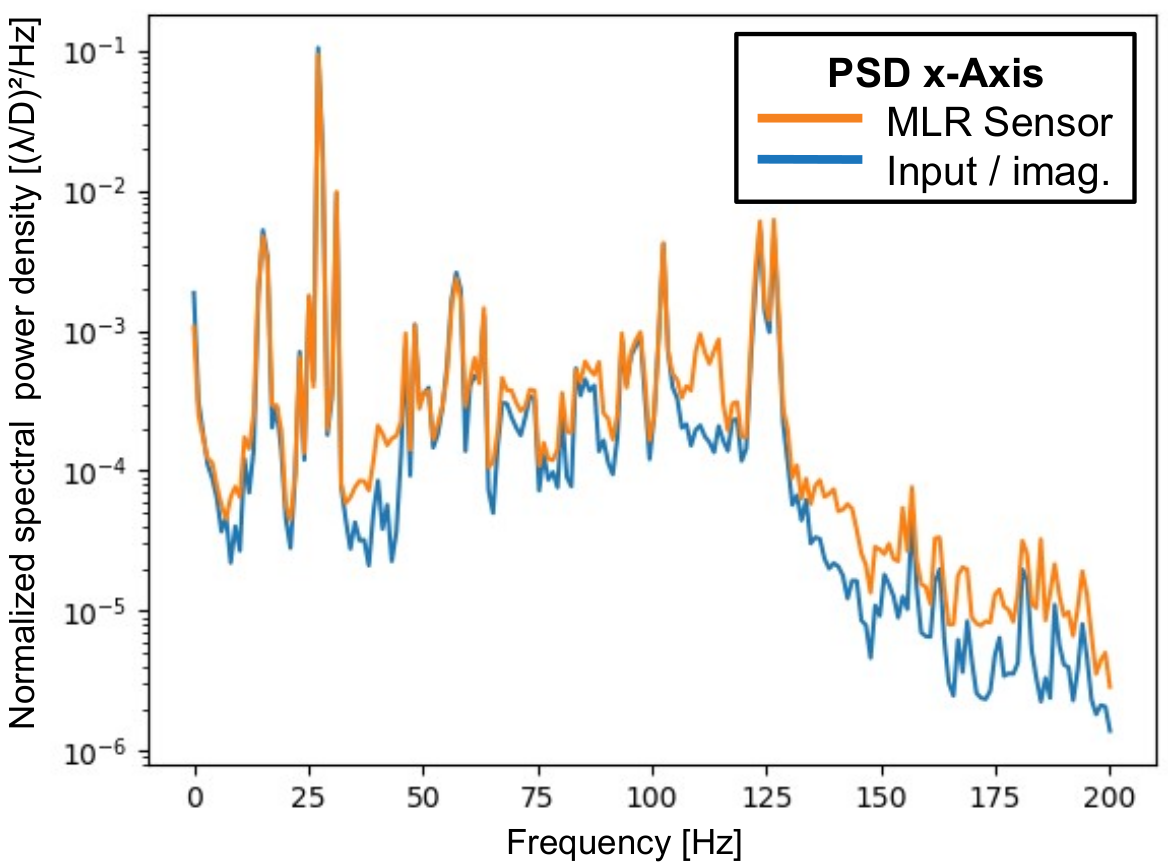}
	\end{center}
	\caption{
  \acf{PSD} of the position in x-direction of introduced vibrations as seen on a separate detector that images the \ac{PSF} (blue) and recovered by the fiber based tip-tilt sensor presented in this work (orange).
  }
	\label{fig:psd-recovery}
\end{figure}

While most of the modeled and designed characteristics are achieved, these discrepancies in coupling efficiency remain and still need to be fully understood.
Furthermore, as seen in Fig.~\ref{fig:response-lab}, the response is not as linear as expected.
This calls for a more complicated reconstruction algorithm.
A simple fitting approach with a sine function was able to recover the measured \ac{PSD} quite accurately.
This is shown in Fig.~\ref{fig:psd-recovery} for reconstruction (orange) of an artificially introduced vibration (blue).
Further calibration correction is being developed to increase accuracy.

\section{Discussion}
\label{sec:discussion}

The setup of the correction system and laboratory results show that the fiber based tip-tilt sensor is capable of sensing aberrations.
Yet, both coupling efficiency of the \ac{SMF} at $58\%$ is less than expected ($67\%$) and coupling into the sensing \acp{MMF} yields considerably less light than modeled ($2.3\%$ compared to $10\%$).
The response also shows a deviation from the predicted linear response, requiring higher order reconstruction algorithms.

When introducing higher order aberrations, the sensor response also shows very characteristic signals.
This can already be used to identify signatures of individual modes and will be further improved to yield \ac{NCP} wavefront data.

Compared to conventional beam stabilization strategies the fiber based tip-tilt sensor presented in this work yields several advantages.
While most techniques direct light off to a separate detector such as a quad-cell \cite{Esposito1997}, the fiber based tip-tilt provides an excellent point of measurement shortly before the focal plane feeding the science fiber.
This ensures that there are no \ac{NCP} aberrations between the science fiber and the sensor, allowing the observer to optimally couple light from the telescope.
Furthermore it is very compact and can easily be integrated into any (existing) instrument, only requiring the space for the fiber itself while the read out optics and electronics can be placed in a remote location.
This can reduce complexity and cost for different applications.
Further research will therefore go into implementing this sensor in small, compact systems and telescopes.
Other advantages are the very predictable vignetting of the light within the system and a wide dynamical range as light is coupled into the sensing fibers even for rather large offsets making it also suitable for coarse (initial) alignment processes.

\section{Conclusion}
\label{sec:conclusion}
\acresetall

In this work we present a fiber based tip-tilt sensor that has been designed to improve \ac{SMF} coupling at the iLocater front end and presented initial laboratory results.
Our novel sensor shows a very distinctive response to a misaligned incoming beam.
Yet, the coupling efficiencies on both the central science \ac{SMF} and the surrounding sensing \acp{MMF} are lower then expected.
Furthermore, an unexpected non-linearity calls for a more complicated reconstruction algorithm.

The sensor is made possible by new exciting technologies such as the two-photon polymerization used for manufacturing the \ac{MLR} for this device.
Only little light is used for sensing if the beam is aligned and only when the beam becomes misaligned more light is refracted into the sensing fibers.
Its advantages are its compact design and sensing at the fiber coupling focal plane, which are not possible with traditional systems, and the potential to sense higher order aberrations.
We plan to test it with realistic on-sky conditions at the iLocater front-end at \acl{LBT} in the near future. Coupled to a suitable \ac{AO} system, this could be an important tool for coupling \acp{SMF} to \ac{ELT} class telescopes.

\acknowledgments

This work was supported by the Deutsche Forschungsgemeinschaft (DFG) through project 326946494, \'Novel Astronomical Instrumentation through photonic Reformatting\'.

This publication makes use of data generated at the Königstuhl Observatory Opto-mechatronics Laboratory (short: KOOL) which is run at the Max-Planck-Institute for Astronomy (MPIA, PI Jörg-Uwe Pott, \url{jpott@mpia.de}) in Heidelberg, Germany. KOOL is a joint project of the MPIA, the Landessternwarte Königstuhl (LSW, Univ. Heidelberg, Co-I Philipp Hottinger), and the Institute for System Dynamics (ISYS, Univ. Stuttgart, Co-I Martin Glück). KOOL is partly supported by the German Federal Ministry of Education and Research (BMBF) via individual project grants.

\bibliography{library}
\bibliographystyle{spiebib} 

\end{document}